\documentclass{PoS}
\usepackage{epsfig}
\usepackage{colordvi}
\usepackage{graphicx}
\usepackage{bm}
\usepackage{multirow}
\usepackage{enumerate}
\reversemarginpar

\def\<{\langle}
\def\>{\rangle}

\newcommand{\text}{\rm}
\def\Tr{{\rm Tr}\,}
\def\tr{{\text tr}\,}
\def\Eq#1{Eq.~(\ref{#1})}

\def\tilde{\widetilde}

\title{Breakdown of large-N reduction in the quenched Eguchi-Kawai
model}

\ShortTitle{Breakdown of large-N reduction in the quenched
Eguchi-Kawai model}

\author{\speaker{Barak Bringoltz}\thanks{This study was supported in
part by the U.S. Department of Energy under Grant
No. DE-FG02-96ER40956.}\\ Department of Physics, University of
Washington, Seattle, WA 98195-1560, USA\\ E-mail:
\email{barak@phys.washington.edu}}

\author{Stephen R.~Sharpe\\ Department of Physics, University of
        Washington, Seattle, WA 98195-1560, USA\\ E-mail:
        \email{sharpe@phys.washington.edu}}

\abstract{We study the validity of the large-N equivalence between
four-dimensional SU(N) lattice gauge theory and its momentum quenched
version---the Quenched Eguchi-Kawai (QEK) model. We have found
strong evidence that this equivalence does not hold in
the weak-coupling regime (and thus in the continuum limit).
This is based on weak-coupling analytic arguments and Monto-Carlo
simulations at intermediate couplings with $20 \le N \le 200$.
Since detailed expositions of our arguments, methods and results
have already appeared in Phys.\ Rev.\ D78:034507 (2008) and Phys.\ Rev.\ D78:074503 (2008), we attempt here to 
give a more intuitive explanation of our results.
The breakdown of reduction that we find is due to a dynamically 
generated correlation between different Euclidean components of the
gauge fields.}

\FullConference{The XXVI International Symposium on Lattice Field
                 Theory\\ July 14-19 2008\\ Williamsburg, Virginia,
                 USA}

\begin{document}

\section{Introduction}

\vspace{-0.35cm}

QCD simplifies in the 't~Hooft limit of a large number of colors, and
as a result it has been a long-standing goal to understand the
properties of the theory in that limit~\cite{largeN}, including on the
lattice~\cite{lattice-reviews}.  In this paper we reconsider an
alternative to conventional large volume simulations, namely the use
of large-$N$ volume reduction to single-site models
\cite{EK,BHN1,TEK,AEK} (see also the related Ref.~\cite{KNN}).
This allows one, in principle, to study very large values $N\sim
O(100-400)$ with modest resources. In this paper we choose to study
one of the variants of the original Eguchi-Kawai (EK) volume
reduction, namely the quenched Eguchi-Kawai (QEK) model. Our
motivation is two-fold: First, it is the only single-site model whose
equivalence with large-$N$ QCD has yet to be thrown in doubt (in
contrast to twisted Eguchi-Kawai model---see
Refs.~\cite{TEKpapers}); Second, we use it as a tool to gain
experience with single-site large-$N$ models, before turning to 
the theory with adjoint fermions~\cite{AEK}. 

Since a detailed discussion of this work,
including a full list of relevant references, 
has already appeared~\cite{QEKpapers}, 
we use this opportunity to give a less
technical and more intuitive presentation.

\vspace{-0.25cm}
\section{A brief review of the EK and QEK models}
\label{review}

\vspace{-0.25cm}

The $U(N)$ EK model of Ref.~\cite{EK} is a matrix model whose
connected correlation functions are expected to have, under certain
assumptions, the same large-$N$ limits as appropriately chosen 
correlation functions in a $U(N)$ pure gauge
theory. It is a specific example of an orbifold 
projection mapping a ``mother'' theory (the gauge theory in our case)
to a ``daughter'' theory (here the matrix model), that,
under certain conditions, becomes
an equivalence in the large-$N$ limit.
This equivalence applies only to
``neutral sectors'' of the two theories,
which consist of correlation functions
invariant under translations in the gauge theory,
and invariant under the $U(1)^d$ ``center'' symmetry
in the matrix model~\cite{KUY} ($d$ is the number of space-time dimensions). 
Obtaining the reduced model from the gauge theory is easy: 
one simply sets to zero all Fourier components
of the gauge fields except for the zero mode and so performs the
mapping $U_\mu(x) \to U_\mu$. Substituting this into the pure gauge
Wilson action gives the partition function of the EK model,
a single-site lattice gauge theory:\footnote{%
The $U(1)^d$ center symmetry is $U_\mu\to e^{i \theta_\mu} U_\mu$.
The gauge symmetry is $U_\mu\to G\; U_\mu\; G^\dagger$ for all $\mu$, 
$G\in U(N)$.}

\begin{equation}
Z_{EK} = \int DU \, \exp(-{S_{EK}})\,, \qquad S_{EK} = - N b
\sum_{\mu<\nu} 2 {\rm Re}\,\Tr(U_\mu U_\nu U_\mu^\dag U_\nu^\dag) \,.
\label{SEK}
\end{equation}
\vspace{-0.25cm}

 The essence of large-$N$ reduction is that vacuum expectation values (VEVs)
 of neutral operators that map into each other have coinciding large-$N$
 limits. For this to hold the vacuum has to be
 invariant under the symmetries defining the projection, for otherwise
 the desired expectation values do not lie in the neutral sectors.
 Indeed the  original EK paper \cite{EK} focused on center-invariant
 VEVs of Wilson loops projected to zero
 momentum. Assuming implicitly that translation invariance is unbroken
 in the gauge theory, while stressing explicitly that the center 
 symmetry must remain intact in both the gauge theory and matrix
 model,
 the authors proved  that these VEVs become equal to corresponding quantities
 in the matrix model at large-$N$.
 (Technically, they showed that  the Dyson-Schwinger equations obeyed by 
 Wilson loops in the two theories coincide.)

Unfortunately, the vacuum of the EK model was shown to
spontaneously break its center symmetry for 
weak enough bare lattice coupling if $d>2$~\cite{BHN1,MK,Okawa1}. 
Consequently, the EK model does not reproduce
the neutral sector of large-$N$ QCD. Initially this was seen in
the weak-coupling limit by analyzing the
effective potential, $V_{\rm eff}$, felt by the eigenvalues of the
link matrices $U_\mu$~\cite{BHN1}. The eigenvalues are attracted, leading to
a ground state with $U_\mu$ proportional to the identity up to a phase,
which spontaneously breaks the center symmetry.
To overcome this they suggested quenching the
eigenvalues, thereby defining a new prescription
for calculating neutral-sector expectation values,
$\<{\cal O}\>_{\rm quenched}$. This prescription has two parts:
\begin{enumerate}
\item 
Calculate $\<{\cal O}\>_p$ for a frozen set of eigenvalues:
${\rm Eig}(U_\mu) = e^{ip^\mu_a}\,;\ a=1,2,\dots,N$. Here $\<,\>_p$
denotes an average with an action obtained from \Eq{SEK} by freezing
the fluctuations in the eigenvalues of $U_\mu$. This means writing
$U_\mu = V_\mu \, e^{ip^\mu}\, V^\dag_\mu$, 
with $p^\mu={\rm diag}(p_1^\mu,\dots,p_N^\mu)$ the frozen eigenvalues
and $V_\mu \in U(N)$, and replacing $DU_\mu$ with $DV_\mu$.
\item 
Average over the $p^a_\mu$ with a measure $d\,\mu(p)$ that
is both $U(1)^d$ invariant and distributes each of the $p_a^\mu$ uniformly
and independently over $[0,2\pi)$ when $N\to\infty$:\\
 $\<{\cal O}\>_{\rm quenched} \equiv \int \, \prod_{a,\mu}
d\mu(p_a^\mu) \, \<{\cal O}\>_p$.
\end{enumerate}
We stress that this prescription differs substantially from the
ordinary way one calculates expectation values in field theory. 
In particular,
for the quenched prescription to coincide with an ordinary field
theory average one would need to use
$d\mu(p) \propto \prod_{a,\mu} d p_a^\mu
e^{-V_{\rm eff}(p)}$~\cite{Migdal}.
This is a highly non-uniform measure over the eigenvalues, as noted
above. By removing the $V_{\rm eff}$ factor, e.g. by using the uniform
measure $d\mu(p) \propto \prod_{a,\mu} d p_a^\mu$,
one expects the center symmetry to remain unbroken.

It is illuminating to understand the quenching prescription in a
different way~\cite{BHN1,Parisi-papers,GK}.
The approach is first formulated for two-index scalar fields.
For example, one can define a mapping between the
large-volume theory with $N\times N$ adjoint fields $\phi(x)$ and a theory of
$N\times N$ matrices $\tilde\phi$: $\phi^{ab}(x) \to
e^{i\sum_\mu (p_a-p_b)^\mu \, x_\mu} \, \tilde \phi^{ab}$. Here $a,b$
are color indices, $x_\mu$ is the space-time coordinate, and
$p_{a}^\mu$ are predefined variables which are referred to as momenta
for a reason that will become clear below. Substituting this mapping into
the field theory action and observables 
yields the matrix model, whose
expectation values depend on the $p_{1,2,\dots,N}^\mu$. 
Next, one is instructed to integrate such expectation values
over $p$. Under certain assumptions, the
result will become equal, when $N\to\infty$,
to the corresponding expectation values in the field theory.

This equality can be shown to any order in perturbation theory
(perturbing in, say, the three-point coupling of the scalar theory)~\cite{GK}.
At large $N$, planar diagrams dominate.
For such diagrams, one finds that, prior to the $p$ integral, 
the perturbative expression in the matrix model at a given order 
coincides with the momentum space {\em integrand} of the corresponding
Feynman diagram in the field theory.
In this correspondence the momentum $q^\mu$ of the field $\phi^{ab}(q)$ 
is given by the difference $q_{ab}^\mu \equiv p_a^\mu-p_b^\mu$
in the matrix model.
The $p$ integration can then be identified as being over the momenta
flowing in the planar diagrams, and one recovers the field theory 
{\em integral}.

Thus the two steps in the quenching prescription can be thought of as
(1) calculating the contribution to an expectation value $\<{\cal
O}\>_p$ of a point $p^\mu$ at a Euclidean Brillouin Zone, and (2)
integrating over the Brillouin zone uniformly to get the full
$\<{\cal O}\>_{\rm quenched}$.  This way of embedding space-time (or
rather its first Brillouin zone) into index space is typical of volume
reduced models and was already noted in 
Refs.~\cite{BHN1,Migdal}.

Performing the mapping in a lattice gauge theory, 
so that $\phi^{ab}(x)$ becomes $U^{ab}_\mu(x)$, 
one obtains a matrix model with an action similar to that in \Eq{SEK},
but with $U_\mu$ replaced by $U_\mu \, e^{ip^\mu}$. 
The momentum factor $e^{ip^\mu}$ can, however, 
be absorbed by a change of variables,
so one ends up back at the problematic EK model.
Various approaches have been used to avoid this problem.
Refs.~\cite{Parisi-papers} restrict their study to $O(N)$
models, while Refs.~\cite{GK,DW} change the measure of the path integral: 
\begin{enumerate}
\item Ref.~\cite{GK}: $\int_{U(N)} DU_\mu \to
\int_{U(N)} DU_\mu\, \int_{U(N)} DV_\mu \, \delta (U_\mu
e^{ip^\mu}-V_\mu \, e^{ip^\mu} V^\dag_\mu )$ (up to $p$-dependent terms), 
thus forcing the eigenvalues of $U_\mu e^{ip^\mu}$ to be $e^{ip^\mu_a}$. 
Intuitively, $U_\mu$ is being forced to fluctuate around unity, 
so that absorbing $p^\mu$ into $U_\mu$ is impossible.
Note that the integral over $V_\mu$ could equally well be over 
the coset $U(N)/U(1)^N$ in which multiplication from the right by
a $U(1)^N$ matrix is divided out.
\item Ref.~\cite{DW}: $\int_{U(N)} DU_\mu \to
\int_{U(N)/U(1)^N} DU_\mu$ (together with a gauge-fixing term). 
Here one restricts the integration regime to the coset 
such that $U_\mu$ cannot absorb the $e^{ip^\mu}$ factor.
\end{enumerate}

The first prescription is, in fact,
identical to the QEK model of Ref.~\cite{BHN1},
as can be seen by using
the $\delta-$functions to perform the $U$ integrations~\cite{GK}. 
The second prescription is different and we call it the DW model.
An obvious advantage of the QEK over the DW model 
is that the ``reduced'' $U(N)$ gauge invariance 
$U_\mu e^{ip^\mu} \to \Omega U_\mu e^{ip^\mu} \Omega^\dag$ 
is realized in the QEK model as $V_\mu \to \Omega V_\mu$. 
By contrast, the DW model has no gauge symmetry:
if one writes $\Omega U_\mu e^{ip^\mu} \Omega^\dag = U'_\mu e^{i p^\mu}$
then, in general, $U'_\mu\notin U(N)/U(1)^N$.
Indeed, the DW model is {\em defined} including
gauge-fixing so as to avoid this problem~\cite{DW}.
This means that a numerical study of the DW model
appears quite nontrivial.

Let us restate in another way why one needs to
change the measure of the path integral over the $U_\mu$.
In the weak coupling regime, $b\to \infty$, configurations close to the
minima of the action dominate.
To avoid the symmetry-breaking of the EK model, 
one wants there to be a single minimum
(up to gauge transformations).
For the action (\ref{SEK}) with $U_\mu$ replaced with $U_\mu e^{ip^\mu}$, 
however, there are multiple minima, occurring when the $U_\mu e^{ip^\mu}$ 
in all directions are arbitrary diagonal matrices
(up to gauge transformations).
By changing the measure as in Refs.~\cite{GK,DW} one
aims to make the only minimum available that at $U_\mu=1$. 
This is indeed what happens, by construction, in the DW model.
One might expect the same for the QEK model: 
the action is minimized when $V_\mu \, e^{ip_\mu} \, V^\dag_\mu \in U(1)^N$
(up to gauge transformations),
and this condition is satisfied by $V_\mu =1$.\footnote{%
Up to right-multiplications by $U(1)^N$ matrices
which we can avoid by restricting $V_\mu$ to $U(N)/U(1)^N$.}
This, however, is not the whole story: as has long been known,
there are other solutions to this equation given by $V_\mu = P_\mu$
with $P_\mu$ a $U(N)/U(1)^N$ matrix that, 
when conjugating $e^{ip_\mu}$, gives rise to
a permutation of the color indices of $p^a_\mu$. 
These minima are nonperturbative 
($P_\mu$ is very far from the unit matrix) 
and we find that they play a central
role in the (in)validity of large-$N$ quenched reduction.

We began this section by describing reduction as an example of the
orbifolding paradigm. 
Once one quenches the daughter theory, however, this paradigm
does not apply in a straightforward way.
This is because VEV's in the daughter QEK model are calculated
in a way that separates the gauge fields into quenched and
unquenched degrees of freedom---a separation that is absent
from the mother gauge theory.
Nonetheless, an argument for quenched reduction has
been given using the
Dyson-Schwinger equations for Wilson loops~\cite{GK}.
These ``loop equations'' are different in the gauge theory and QEK model,
but the differences are proportional to quenched expectation values
of center-invariant quantities composed of center non-invariant Wilson loops. 
An example of such a quantity is $\langle|W_{C}|^2\rangle_{\rm quenched}$, 
where $W_C$ is center non-invariant.
Reduction can hold only if such quantities vanish as $N\to\infty$.
They will vanish if quenched expectation values factorize, for then 
$\langle|W_{C}|^2\rangle_{\rm quenched}
= |\langle W_{C}\rangle_{\rm quenched}|^2+ O(1/N)$, 
and the enforcement of center-symmetry in the QEK implies that
$|\langle W_{C}\rangle_{\rm quenched}|=0$.
The crucial question is thus whether factorization holds
for quenched expectation values.

\vspace{-0.35cm}
\section{(In)Validity of large-$N$ quenched reduction }

In this section we discuss analytic,
weak-coupling considerations that extend those presented in
Refs.\cite{BHN1} and \cite{MK} to include the effects of the multiple
classical minima which, as discussed above, come in the form of
permutations of the $p_a^\mu$. 
To be precise, the minima are $U^{\rm min}_\mu e^{ip^\mu} = V^{\rm
min}_\mu \, e^{ip^\mu}\, V^{\dag,{\rm min}}_\mu = e^{i\tilde
p^\mu}$, with $\tilde p_a^\mu = p_{\sigma(a)}^\mu$ and
$\sigma(a)$ one of the $N!$ permutations of the indices $a$.
Here we describe how this classical
degeneracy is removed by quantum fluctuations, 
with the resulting ground state corresponding to a certain 
permutation which is unrelated to the original choice of $p_a^\mu$. 
This, by itself, is a strong
indication for the breakdown of quenched large-$N$ reduction, since it
invalidates the idea that for each value of $p^\mu$, the value of
$\<{\cal O}\>_p$ is the contribution of a point corresponding to
$p^\mu$ in the Brillouin zone.

To show how quantum fluctuations remove the classical degeneracy we
minimize the effective potential $V_{\rm eff}$ 
over the space of all permutations of the color indices 
of the input momenta $p_a^\mu$
(with permutations in different directions being independent).
For brevity, we consider here the case where the $N$ momenta in each direction,
$p_a^\mu$, are always a permutation of the ``clock'' momenta 
$p_a^{\rm clock} \equiv 2\pi a/N$. 
The situation for other distributions of momenta
is similar, as discussed in Refs.~\cite{QEKpapers}. 
For clock momenta we find that, for $d\ge 3$,
the minimum of $V_{\rm eff}$ occurs for permutations in which
the momenta in all directions are ordered similarly. 
In equations, the statement is that
$\tilde p_a^\mu - \tilde p_a^\nu = \alpha^{\mu\nu}$ for all $a$,
and we refer to this as ``locking'' of the momenta.
There is thus an overall shift between different Euclidean directions, 
whose value is an arbitrary clock momentum that can depend on $\mu$ and $\nu$,
but which is independent of the color index. 
This implies that the difference $q_{ab}^\mu=\tilde p_a^\mu-\tilde p_b^\mu$, 
which we recall plays the role of the gluon momentum in Feynman diagrams, 
{\em is independent of $\mu$}.
This in turn means that instead of integrating over the full Brillouin zone
(as would be the case if the $p^\mu$ in different directions were
independent) one is effectively integrating along the diagonal of the
Brillouin zone, i.e. over a one-dimensional line!
 
\begin{figure}[htb]
\centerline{
\begin{tabular}{cc}
\parbox{7cm}{
\includegraphics[width=7cm]{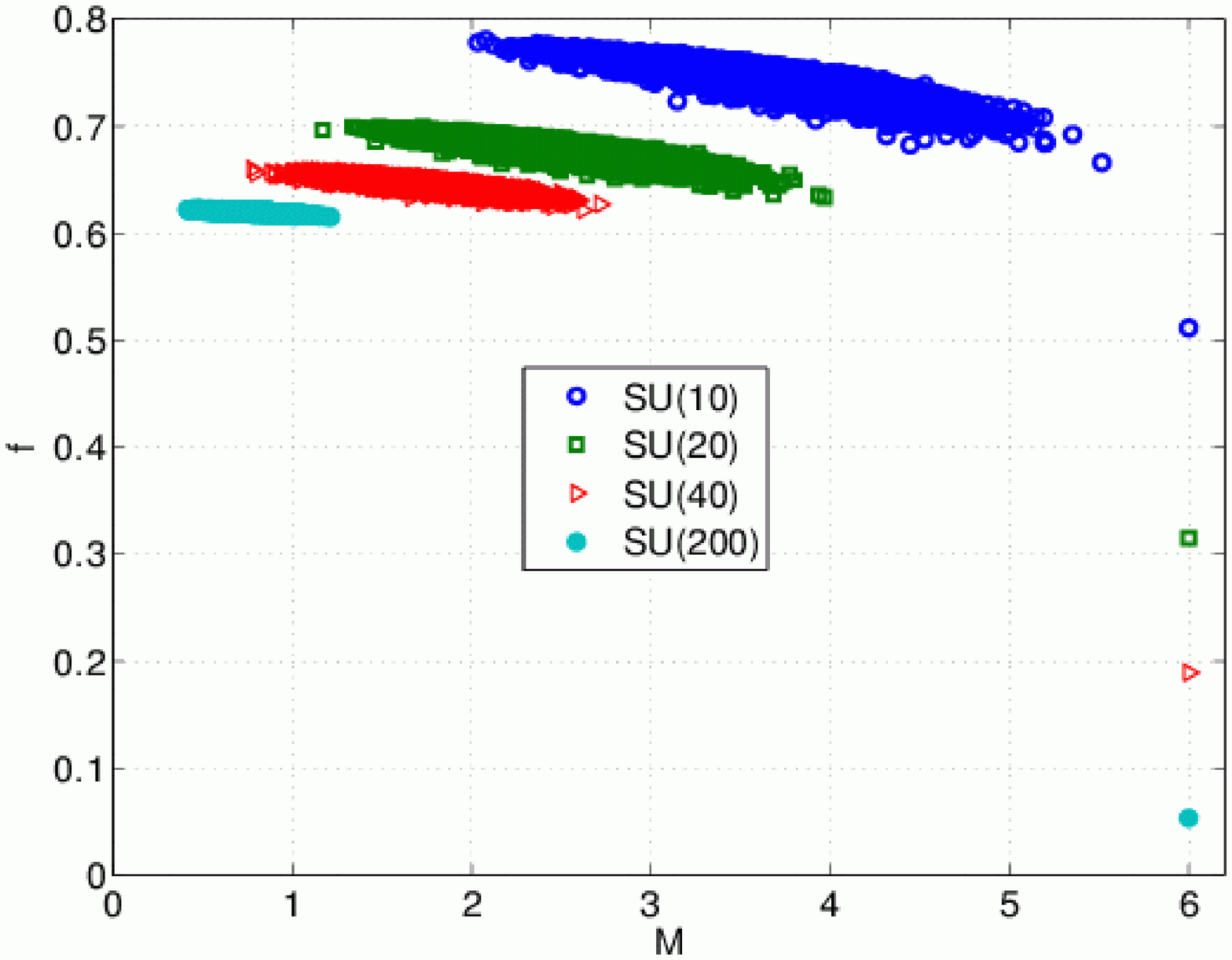} }
\quad & \quad
\parbox{8cm}{\includegraphics[width=8cm]{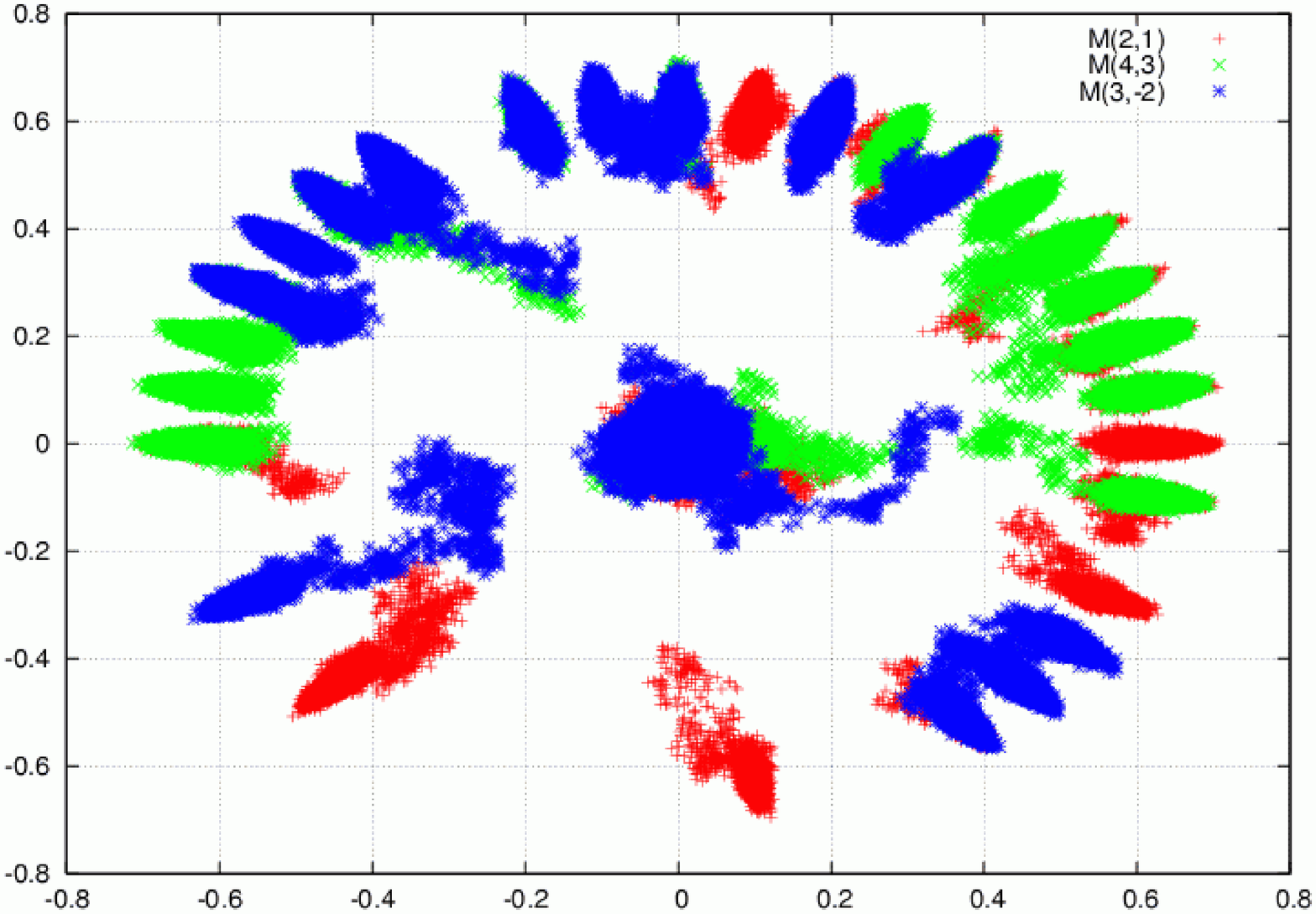}}
\end{tabular}
}

\caption{\underline{Right panel:} The dependence of $V_{\rm eff}/N^2$ on
the combined order parameter $M$ for random permutations of the clock
momenta and for various values of $N$. \underline{Left panel:} Scatter
plot of the values of three chosen $M_{\mu\nu}$'s in a run for $N=40$
and `clock' momenta. We refer to \cite{QEKpapers} for technical
details.}
\label{fig1}
\end{figure}

It is useful to have order parameters that can detect such ``locking'',
and a simple choice is the center-non-invariant reduced Wilson loops
$M_{\mu,\nu}\equiv \frac1N \, \tr \, U_\mu \, U_\nu$ and
$M_{\mu,-\nu}\equiv \frac1N \, \tr \, U_\mu \, U^\dag_\nu$.\footnote{%
We note that with clock momenta one always has $\tr\,U_\mu = 0$.}
Indeed, when the locked ordering takes place one finds that 
$M_{\mu,-\nu} = e^{i\alpha^{\mu\nu}}$, while $M_{\mu,\nu}=0$. 
By contrast, if no ordering takes place and all $p_a^\mu$ are random 
permutations of $p^a_{\rm clock}$, 
then $M_{\mu,\pm\nu}\sim O(1/N)$. In Fig.~\ref{fig1} we show the
dependence, in $d=4$, of $V_{\rm eff}$ on the
combined order parameter $M = \frac12 \sum_\mu \,\left( |M_{\mu,\nu}|
+ |M_{\mu,-\nu}|\right)$, where we
have used a Monte-Carlo (MC) to sample permutations.
One sees that $V_{\rm eff}$
has a minimum at the ``fully locked'' ground state with $M=6$,
with states having little locking ($M\ll1$) having a free energy of
$O(N^2)$ higher.

We now come to a crucial observation: if locked ordering occurs
then large-$N$ factorization breaks down in the quenched theory.
To see this, recall that the quenched prescription instructs us
to integrate over the $p_a^\mu$ in a center-symmetry invariant way.
Performing a center transformation on the $p_a^\mu$
($p_a^\mu\to p_a^\mu+\theta_\mu$) will lead to
a center-transformed locked vacuum, since the action is center invariant.
In this transformed vacuum, 
$\alpha^{\mu\nu}\to \alpha^{\mu\mu}+\theta_\mu-\theta_\nu$,
so the phase of the $M_{\mu,\nu}$ is shifted.
Integrating over the $\theta_\mu$, which is part of the
integration over the $p_a^\mu$, will thus lead to the vanishing
of the order parameters, e.g. in the example above,
$\<M_{\mu,-\nu}\>_{\rm
quenched}\simeq\int_0^{2\pi} d\alpha^{\mu\nu} \, e^{i\alpha^{\mu\nu}}=0$. 
By contrast, if one calculates
$\<|M_{\mu,-\nu}|^2\>_{\rm quenched}$,
the phase $\alpha^{\mu\nu}$ cancels,
and so $\<|M_{\mu,-\nu}|^2\>_{\rm quenched}\sim O(1) \neq
|\< M_{\mu\nu}\>_{\rm quenched}|^2$. 
This invalidates large-$N$ factorization in the quenched theory,
which, as noted above, implies that the gauge theory and QEK model
have different loop equations, and so reduction fails.

The argument for locked ordering described above was perturbative.
To check whether it occurs beyond
perturbation theory we performed a detailed numerical study of the QEK
model at intermediate and strong couplings using MC
techniques for $20 \le N\le 200$. We do not discuss any
particulars of these calculations here, referring instead to
Refs.~\cite{QEKpapers} (the second reference includes an appendix
describing the update algorithms we used, which
are nontrivial since the model is {\em quartic} in the updated
fields).

In our numerical studies we measure the order parameters $M_{\mu\nu}$
in the intermediate coupling regime,
and find clear evidence for correlations in the eigenvalue
orderings. For example, in the right panel of Fig.~\ref{fig1} we show
a typical scatter plot of data obtained for $M_{2,1}$, $M_{4,3}$, and
$M_{3,-2}$. This combines results from 20 independent input
choices of the $p_a^\mu$, all being random permutations of $p_a^{\rm
clock}$. What we see from the figure is that each of these $20$
MC simulations settled into a different vacuum
characterized by a distinct value for the phase $\alpha^{\mu\nu}$
of $M_{\mu\nu}$. This is precisely what one expects when locking
of the eigenvalue ordering occurs.

We have also obtained direct evidence for the breakdown of
large-$N$ reduction: there are large discrepancies between the
plaquette values and the structure of the phase diagram of the QEK
model and the large-$N$ gauge theory~\cite{QEKpapers}.  
We have checked that these conclusions are
insensitive to the precise form of the quenched eigenvalue
distribution, $d\mu(p)$, and to the way we perform the quenched
average. We also considered values of $N$ up to $200$ to look for a
late onset of $1/N$ behavior, but find none. Therefore we 
conclude that the momentum quenched large-$N$ reduction of $SU(N)$
lattice gauge theories fails in the weak-coupling regime
and thus in the continuum limit.

\vspace{-0.35cm}
\section{Possible future directions}

\vspace{-0.25cm}

Our result, together with those of Refs.~\cite{TEKpapers}, imply that
the only known single-site models that can
reproduce the properties of QCD at large-$N$ are the
following: (I) The ``deformed'' Eguchi Kawai of Ref.~\cite{DEK}; 
(II) the single-site obtained by adding adjoint fermions to the
Eguchi-Kawai model \cite{AEK}; and (III) the momentum quenched
model of DW~\cite{DW}. For a brief discussion on the first two we refer
to Refs.~\cite{QEKpapers}. As far as we know, the third alternative has
never been explored nonperturbatively; its obvious advantage over the
standard type of momentum quenching is that its path integral has a
unique minimum in weak coupling, with no difficulties related to
permutations.
As mentioned above, however, this model involves
gauge fixing and so it is clearly more involved to
study numerically. 
We leave the exploration of all these single-site models to future
studies.

\end{document}